\documentclass[twocolumn, eqsecnum,showpacs,preprintnumbers,amssymb,aps,prl]{revtex4}
\usepackage{graphicx}
\usepackage{dcolumn}
\usepackage{bm}

\begin{document}

\title{Investigations of Ra$^+$ properties to test possibilities of new optical frequency standards}
\vspace{0.3cm}

\author{$^a$B. K. Sahoo \protect \footnote[2] {E-mail: bijaya@mpipks-dresden.mpg.de}, $^b$B. P. Das, $^b$R. K. Chaudhuri, $^c$D. Mukherjee, $^d$R. G. E. Timmermans and $^d$K. Jungmann \\
{\it $^a$Max Planck Institute for the Physics of Complex Systems, D-01187 Dresden, Germany}\\
{\it $^b$Non-Accelerator Particle Physics Group, Indian Institute of Astrophysics, Bangalore-34, India}\\
{\it $^c$Department of Physical Chemistry, IACS, Kolkata-700 032, India}\\
{\it $^d$KVI, University of Groningen, NL-9747 AA Groningen, The Netherlands}}

\date{Received date; Accepted date}

\vskip1.0cm
\begin{abstract}
\noindent
The present work tests the suitability of the narrow transitions $7s \ ^2S_{1/2} \rightarrow 6d \ ^2D_{3/2}$ and $7s \ ^2S_{1/2} \rightarrow 6d \ ^2D_{5/2}$ 
in Ra$^+$ for optical frequency standard studies. Our calculations of the lifetimes of
the metastable $6d$ states using the relativistic coupled-cluster theory suggest
 that they are sufficiently long for Ra$^+$ to be considered as a potential 
candidate for an atomic clock. This is further corroborated by our studies of the hyperfine interactions, dipole and quadrupole polarizabilities and quadrupole
moments of the appropriate states of this system.
\end{abstract} 
\maketitle

Accurate time and frequency measurement is crucial for the advance of science 
and technology in many fields. This leads to a number of searches to find
candidates for optical frequency standards. The current frequency standard is
based on the ground state hyperfine transition in atomic cesium and has a 
quality factor (Q) of 
$10^{15}$ \cite{essen}. Atomic spectral lines with high Q are generally
interesting for standards, however good control over systematic line shifts
will be essential. As a result of the remarkable advances in the field of ion 
trapping and laser cooling, single ions like Hg$^+$ \cite{oskay}, In$^+$ 
\cite{becker}, Ca$^+$ \cite{champenois}, Sr$^+$ \cite{barwood}, Yb$^+$
\cite{blythe}, Cd$^+$ \cite{tanaka} and Ba$^+$ \cite{sherman} are particularly
interesting as they can be localized using their electric charge rather than 
light forces, which is necessary for atom trapping. Very accurate
measurements have been performed on Hg$^+$ and Sr$^+$, where Q exceeds 
$10^{17}$. Some of the major systematic errors associated with the clock
frequency are the Stark
effect, Zeeman effect and quadrupole shifts due to stray electric 
fields in the ion trap \cite{itano}. These errors can be estimated 
from high precision theoretical studies of
hyperfine structure constants, polarizabilities and quadrupole moments of
the appropriate atomic states. Indeed, studies of these quantities are also  
essential for parity non-conservation (PNC) studies \cite{fortson,bijaya0}.
Some of the above mentioned errors can be eliminated by considering
the clock transition between suitable hyperfine states \cite{oskay}.

\begin{figure}[h]
\includegraphics[width=6.5cm]{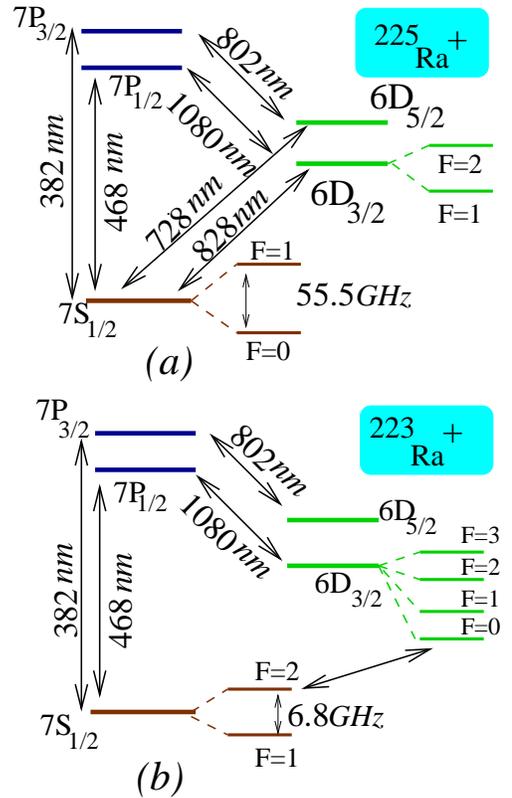}
\caption{(color online) Schematic diagram of energy levels of Ra$^+$ with
transitions for possible optical frequency standards.}
\label{fig1}
\end{figure}
An experiment is in progress at KVI to search for an suitable 
optical frequency standard by measuring the frequency of either $7s \ ^2S_{1/2} \rightarrow 6d \ ^2D_{3/2}$ or $7s \ ^2S_{1/2} \rightarrow 6d \ ^2D_{5/2}$
transitions in Ra$^+$. A similar experiment is also being planned at IACS
\cite{manas}. In this paper, we report our theoretical studies on the
feasibility of these transitions for the optical frequency studies in Ra$^+$.
In the case of Ba$^+$
it has been pointed out that PNC and optical frequency standards experiments
share many features in common \cite{sherman}. The techniques used in the Ba$^+$
experiments can be extended to Ra$^+$ as the electronic structures of the two
ions are similar. However, Ra$^+$ has one important advantage: the low lying 
transition wavelengths (see Fig. \ref{fig1}({\it a})) of this ion are in the 
optical regime making them more easily accessible than their counterparts in 
Ba$^+$. Although, it appears that these wavelengths can be measured very precisely
using modern spectroscopic techniques, it is however necessary to find out
which transition is the most suitable for optical frequency 
standard. This can be decided by the experimentalists from the knowledge
of different physical quantities of the involved states that can be used to 
control the sources of error.

 First of all, one must determine which of the isotopes of Ra$^+$ merit 
consideration for optical clock studies. In this context, it is worthwhile 
to note that only
$^{223}$Ra and $^{225}$Ra have half-lives of a few days ($\sim$ 10days)
and these isotopes are therefore obvious choices. However, they 
have different nuclear spins ($I$s); the former has $I=3/2$ whereas the latter
has $I=1/2$ and this results in different hyperfine splittings. One has to take 
into account the various systematic errors while considering both these isotopes. It is  
possible to eliminate the quadrupole Stark shift by
considering the transition between the hyperfine ($F$) states such as 
$|6s(J=1/2),I=3/2;F=2\rangle \rightarrow |5d(J=3/2),I=3/2;F=0\rangle$ 
transition (see Fig. \ref{fig1}({\it b})) for the frequency 
standard although knowledge of the hyperfine structure constants
and the polarizabilities are still required for these experiments. It is necessary 
to study the hyperfine structure constants, lifetimes and a few other spectroscopic quantities for the $7s$
and $6d$ states of this system in order to assess the suitability of the proposed clock transitions. 

Electron correlation and relativistic effects must be treated accurately for 
Ra$^+$.
 Relativistic coupled-cluster (RCC) theory; a size-consistent, 
 size-extensive and an all order perturbation method is well suited for this purpose \cite{szabo}. It 
has been successfully applied to determine accurately certain ground and excited states properties of  
Sr$^+$ \cite{bijaya1} and Ba$^+$ \cite{bijaya2}. We employ the same method
in the present study to obtain accurate results for Ra$^+$. The  
presence of the non-linear terms in this method makes it challenging to obtain the 
ground and excited state wave functions for a large system like
Ra$^+$. We had observed earlier that these effects are important
for accurate studies \cite{bijaya3} in other heavy systems.  
In order to obtain the wave functions for Ra$^+$, we solve the RCC equations
considering single, double and leading triple excitations (CCSD(T) method). 
This involves the determination of $10^7$ cluster amplitudes self-consistently.
This is one of the largest 
computations to date for obtaining the wave functions of an atomic system.

The starting point of our work is the relativistic generalization of
the valence universal coupled-cluster (CC) theory introduced by Mukherjee
et al. \cite{mukh} which was put later in a more compact form by Lindgren 
\cite{lind}. In this approach, the atomic wave function $|\Psi_v \rangle$ for 
a single valence ($v$) open-shell system is expressed as
\begin{eqnarray}
|\Psi_v \rangle = e^T \{1+S_v\} |\Phi_v \rangle ,
\label{eqn1}
\end{eqnarray} 
where $|\Phi_v \rangle$ 
is the reference state constructed out of the Dirac-Fock (DF) orbitals of
the closed-shell system ($|\Phi_0 \rangle$) by appending the valence electron 
orbital. Here $T$ and $S_v$ are the excitation operators from the core and 
valence-core sectors (for example see \cite{bijaya1,bijaya3} 
for the second quantization representations of these operators and equations to 
obtain their amplitudes). The single particle orbitals in the present calculations are linear combinations of 
Gaussian type functions  
 \cite{rajat}.

The transition matrix element of a hermitian operator ($O$) corresponding to the initial state 
$| \Psi_i \rangle$ and the final state $| \Psi_f \rangle$ can be expressed 
using the RCC method as 
\begin{eqnarray}
\langle O \rangle_{if} &=& \frac {\langle \Psi_i | O | \Psi_f \rangle} {\sqrt{ \langle \Psi_i | \Psi_i \rangle \langle \Psi_f|\Psi_f \rangle }} \nonumber \\
&=& \frac {\langle \Phi_i |\{1+S_i^{\dagger}\} \overline{O} \{1 +S_f\} | \Phi_f \rangle } {\sqrt{ N_i N_f}} , \ \ \ \ \
\label{eqn2}
\end{eqnarray}
where we define $\overline{O}=e^{T^{\dagger}} O e^T$ and $N_v = \langle \Phi_v |e^{T^{\dagger}}  e^T + S_v^{\dagger} e^{T^{\dagger}} e^T S_v | \Phi_v \rangle$ 
for the valence electron $v$. We calculate the above expression using the 
procedure followed in the earlier works \cite{bijaya1,bijaya2,bijaya3}. The
expectation values are determined by considering the special condition $i=f$.

{\it Lifetimes of the $6d$ states}: It is necessary to know the lifetimes of the
 $6d$ metastable states to understand how reliably the proposed experiments can be
performed in that time period. The lifetimes (in second (s)) of these states 
can be determined from the inverse of the total transition probabilities ($A$). The 
net transition probabilities (in s$^{-1}$) of the $6d$ states are given by
\begin{eqnarray}
A_{6 d5/2} &=& A^{\text{E2}}_{6 d5/2 \rightarrow 7 s1/2} + A^{\text{E2}}_{6 d5/2 \rightarrow 6
 d3/2} + A^{\text{M1}}_{6 d5/2 \rightarrow 6 d3/2}, \nonumber \\
A_{6 d3/2} &=& A^{\text{E2}}_{6 d3/2 \rightarrow 7 s1/2}  + A^{\text{M1}}_{6 d3/2 \rightarrow
7 s1/2} ,
\label{eqn3}
\end{eqnarray}
where 
\begin{eqnarray}
A^{\text{E2}}_{f \rightarrow i} &=& \frac {1.11995 \times 10^{18} }{(2j_f+1) \lambda^5} S^{\text{E2}}_{f \rightarrow i} \\
A^{\text{M1}}_{f \rightarrow i} &=& \frac {2.69735 \times 10^{13} }{(2j_f+1) \lambda^3} S^{\text{M1}}_{f \rightarrow i},
\label{eqn3}
\end{eqnarray}
where $S_{f \rightarrow i} = |O_{fi}|^2$ and $\lambda$ (in \AA) are the transition line strength for the
operator $O$ (in atomic unit (au)) and wavelength, respectively.
These quantities depend on both the transition amplitudes and wavelengths,
and they can be calculated using a single {\it ab 
initio} method. However, we use experimental wavelengths \cite{moore}
to reduce the errors in the determination of the lifetimes.
\begin{table}[h]
\begin{ruledtabular}
\begin{center}
\begin{tabular}{lccccc}
Transition & \multicolumn{2}{c}{$O^{\text{E2}}_{f \rightarrow i}$} & $O^{\text{M1}}_{f \rightarrow i}$ & Lifetime \\
states  & \multicolumn{2}{c}{(au)}  & (au) & (s) \\
${f \rightarrow i}$ & {\it Length} & {\it Velocity} &  & \\      
\hline
     &  &  & & \\
$|6d_{3/2}\rangle \rightarrow |7s_{1/2}\rangle$ & 14.87(7) & 14.77(22) & 0.0024(2) & 0.893(4) \\
     &  &  & & \\
$|6d_{5/2}\rangle \rightarrow |7s_{1/2}\rangle$ & 19.04(5) & 19.87(1.0) &  & 0.301(3) \\
$+|6d_{5/2}\rangle \rightarrow |6d_{3/2}\rangle$ & 8.80(4) & 10.5(2.5) & 1.546(1) & 0.297(4) \\
\end{tabular}
\end{center}
\end{ruledtabular}
\caption{Transition amplitudes (in au) due to M1 and E2 transitions in both
length and velocity gauges. Length gauge results of the E2 amplitudes along 
with M1 amplitudes are considered for the determination of lifetimes.}
\label{tab1}
\end{table}

Since there are no experimental or theoretical predictions of the lifetimes
of the $6d$ states, we calculate the E2 transition amplitudes using
both the length and velocity gauges in order to assess the numerical accuracies of 
the results. These results are given in Table \ref{tab1} along with the
M1 transition amplitudes and the lifetimes of the $6d$ states. We have used the 
E2 amplitudes in the length gauge as it converges faster than the corresponding 
values in the velocity gauge.
 The errors are estimated from the discrepancies of the 
results obtained with different choices of bases.

Using the RCC method, we find that due to the enhanced role of electron correlation, core polarization effects in particular, the M1 transition amplitude for the 
$|6d_{3/2}\rangle \rightarrow |7s_{1/2}\rangle$ transition is 0.0024(2)$ea_0$ 
where the DF value is $\sim 10^{-5}ea_0$. From the calculated E2 amplitudes
of this transition, we obtain the lifetime of the $6d_{3/2}$ state as $0.627(4)$s.
Inclusion of the above M1 transition probability changes its value to 
$0.893(4)$s; this
change is around 30\% of the total result and this finding is different from 
our earlier studies on similar states of other alkaline earth metal ions
\cite{bijaya4}. However, like the other $d_{5/2}$ states in those systems, the 
lifetime of the $6d_{5/2}$ state reduces from 0.301s to 0.297s after 
including the contribution of the M1 transition probability in the $|6d_{5/2}\rangle \rightarrow |6d_{3/2}\rangle$ transition.

{\it Quadrupole moments of the $6d$ states:} In order to estimate the error in 
the frequency of the clock transition arising from quadratic Stark shifts, 
it is necessary to know the quadrupole moments of the relevant states.
The quadrupole moment of a valence state ($v$) 
is given by
\begin{eqnarray}
\Theta(v) = \langle \Psi_v|O^{\text{E2}}|\Psi_v \rangle,
\label{eqn4}
\end{eqnarray}
where $O^{\text{E2}}$ is the E2 transition operator. We divide the the above 
expression into three parts as follows
\begin{eqnarray}
\Theta(v) = \Theta_{DF}(v)+ \Theta_{cv}(v) + \Theta_v(v) .
\label{eqn5}
\end{eqnarray}
Here $\Theta_{DF}$, $\Theta_{cv}$ and $\Theta_v$ are the DF, core-valence
and valence electron correlation effects. In Table \ref{tab2}, we
present these contributions for the $6d_{3/2}$ and $6d_{5/2}$ states. In this 
table, the difference between the total RCC result and the sum of all the above
three contributions is due to the normalization of the wave functions.
The quadrupole moment of the $7s$ state is clearly zero as the quadrupole moment operator is of 
rank two. Therefore, we determine these quantities only for the $6d$ states.

\begin{table}[h]
\begin{ruledtabular}
\begin{center}
\begin{tabular}{lcccc}
State & $\Theta_{DF}$ & $\Theta_{cv}$  & $\Theta_{v}$ & $\Theta$ \\
\hline
    &  &  & & \\
$6d_{3/2}$ & 3.48 & $-$0.01 & $-$0.51 & 2.90(2) \\ 
$6d_{5/2}$ & 5.19 & $-$0.02 & $-$0.65 & 4.45(9) \\ 
\end{tabular}
\end{center}
\end{ruledtabular}
\caption{Quadrupole moments of atomic states in au.}
\label{tab2}
\end{table}
 As given in Table \ref{tab2}, the dominant contribution comes from $\Theta_{DF}$ followed by $\Theta_v$, 
which contains core-polarization and pair-correlation effects to all orders,
make significant contributions as in Sr$^+$ \cite{bijaya1} and Ba$^+$ \cite{bijaya2}. We 
have followed the same procedure as in the lifetime calculations to estimate 
errors in these results.

{\it Polarizabilities:} We determine the dipole polarizabilities for $7s$ and
$6d$ states and quadrupole polarizability of the $7s$ state for our
study. The static 
($\alpha_0^1(J_v)$) and tensor dipole ($\alpha_2^1(J_v)$) polarizabilities 
for the valence $v$ state with angular momentum $J_v$ are given by
\begin{eqnarray}
\alpha_0^1(v) &=& - 4 \sum_{k \ne v} \frac {|\langle J_v|D|J_k \rangle|^2}{E_v - E_k}
\label{eqn6} \\
\text{and} && \nonumber \\
\alpha_2^1(v) &=& 4\sqrt{ \frac {30 j_v (2j_v -1 )(2j_v + 1)}{(j_v+1)(2j_v+3)}}  \sum_{k \ne v} (-1)^{J_v+J_k+1} \nonumber \\
&& \left \{ \matrix { J_v & 1 & J_k \cr 1 & J_v & 2 \cr } \right \} \frac {|\langle J_v|D|J_k \rangle|^2}{E_v - E_k},
\label{eqn7}
\end{eqnarray}
respectively, where $D$ is the E1 operator. Similarly, the static quadrupole
polarizability ($\alpha_0^2(v)$) is given by
\begin{eqnarray}
\alpha_0^2(v) &=& - 4 \sum_{k \ne v} \frac {|\langle J_v| O^{\text{E2}} |J_k \rangle|^2}{E_v - E_k} .
\label{eqn8}
\end{eqnarray}
We have used the 
sum-over-states approach and experimental energies to reduce the errors in the 
calculations; the calculated energies were used obtained from the RCC method 
where the experimental energies were not available.

We express generally the polarizabilities as
\begin{eqnarray}
\alpha(v) &=& \alpha_{DF}(v)+ \alpha_c(v) + \alpha_{cv}(v) + \alpha_v(v) ,
\label{eqn9}
\end{eqnarray}
where each term is defined similar to the corresponding terms of the quadrupole
moment expression given in Eq. (\ref{eqn5}) except for $\alpha_c$ which is the pure core orbital 
contribution. We calculate $\alpha_v$ contributions from the calculated 
intermediate states
using the RCC method. However, $\alpha_c$ and $\alpha_{cv}$ are calculated
using the second order many-body perturbation theory (MBPT(2)), where the  
residual Coulomb interaction and E1/E2 operators are treated as perturbation.
All these results are tabulated in Table \ref{tab3}.
\begin{table}[h]
\begin{ruledtabular}
\begin{center}
\begin{tabular}{lccccccc}
 & State & $\alpha_{DF}$ & $\alpha_c$  & $\alpha_{cv}$ & $\alpha_v$ & $\alpha_{t}$ & $\alpha$s \\
\hline
   &  & &  & & & & \\
$\alpha_0^1$ & $7s_{1/2}$ & 99.39 & 14.57 & $-$1.54 & $-$4.34 & $-$0.02 & 107.86 \\ 
& $6d_{3/2}$ & 926.32 & 14.57 & $-$1.31 & 39.05 & 0.03 & 978.66 \\ 
& $6d_{5/2}$ & 1208.13 & 14.57 & $-$7.53 & 77.18 & 0.03 & 1292.38 \\ 
   &  & & & & & \\
$\alpha_2^1$& $6d_{3/2}$ & $-$220.67 & 1.21 & 0.76 & 65.38 & $-$0.01 & $-$153.12\\ 
& $6d_{5/2}$ & $-$168.53 & 1.21 & 1.27 & 16.78 & $-$0.01 & $-$149.28 \\ 
   &  & &  & & & \\
$\alpha_0^2$ & $7s_{1/2}$ & 2920.83 & 56.57 & 0.34 & $-$436.04 & 5.50 & 2547.20 \\ 
\end{tabular}
\end{center}
\end{ruledtabular}
\caption{Dipole and quadrupole polarizabilities in au.}
\label{tab3}
\end{table}

We have obtained up to $10s$, $10p$, $10d$, $9f$ and $9g$ low-lying states
using the RCC 
method to calculate the above quantities. Contributions from other higher
states are accounted for using MBPT(2). They are just given as tail contributions
($\alpha_t$) in the table. Using the expression
\begin{eqnarray}
\alpha_{0,6d}^2(7s) &=& - \sum_{k=6d_{3/2},6d_{5/2}} \frac {|\langle \Psi_{7s}|O^{\text{E2}}|\Psi_k\rangle|^2}{E_{6s} - E_k} ,
\label{eqn10}
\end{eqnarray}
we obtain  $\alpha_2^0(7s)=1037(7)a_0^5$ along with the corresponding $\alpha_c$
 contribution. This is usually necessary for the lifetime measurements of the 
$6d$ states.
\begin{table}[h]
\begin{ruledtabular}
\begin{center}
\begin{tabular}{lccc}
  & $7s_{1/2}$ & $6d_{3/2}$ & $6d_{5/2}$ \\ 
\hline
   &  & &  \\
$7p_{1/2}$ & 3.28 & 3.64 &      \\
$8p_{1/2}$ & 0.04 & 0.07 &      \\
$7p_{3/2}$ & 4.54 & 1.54 & 4.92 \\
$8p_{3/2}$ & 0.50 & 0.15 & 0.40 \\
$5f_{5/2}$ &      & 4.47 & 1.31 \\
$6f_{5/2}$ &      & 0.86 & 0.21 \\
$5f_{7/2}$ &      &      & 6.21 \\
$6f_{7/2}$ &      &      & 1.08 \\
\end{tabular}
\end{center}
\end{ruledtabular}
\caption{Important reduced E1 matrix elements in au used to determine the dipole polarizabilities.}
\label{tab3d}
\end{table}

In Table \ref{tab3d}, we present the important reduced E1 matrix elements
which are used in the determination of dipole polarizabilities.
These results are in reasonable agreement with those of Dzuba et al. which are
calculated using another many-body approach \cite{dzuba}.

{\it Hyperfine structure constants:} Studies of these constants are important
to investigate the underlying physics of the wave functions in the nuclear
region, especially to estimate the errors of the PNC matrix elements 
\cite{bijaya6}. The magnetic dipole ($A_h$) and electric quadrupole ($B_h$)
hyperfine structure constants of the valence $v$ state with angular momentum 
$J_v$ are given by
\begin{eqnarray}
&& A_h(v) = \frac{\mu_N g_I}{J_v} \langle \Psi_v |\text{T}^{(1)}| \Psi_v\rangle 
\label{eqn11} \\
\text{and} && \nonumber \\
&& B_h(v) = 2e Q_N \langle \Psi_v| \text{T}^{(2)}|\Psi_v\rangle, 
\label{eqn12}
\end{eqnarray}
respectively. In the above expressions, $\mu_N$, $g_I$ and $Q_N$ are the nuclear
magnetic moment, gyromagnetic ratio and quadrupole moment, respectively. 
Explicit expressions and the single particle matrix elements of $\text{T}^{(1)}$
and $\text{T}^{(2)}$ are given in \cite{cheng}. We have used $g_I=0.18067$ 
\cite{arnold} and $Q_N=1.254$ \cite{neu} for $^{223}$Ra and $g_I=-1.4676$ 
\cite{arnold} for $^{225}$Ra in these calculations.
\begin{table}[h]
\begin{ruledtabular}
\begin{center}
\begin{tabular}{lc|c|cc|cc}
    & & $7s_{1/2}$ & \multicolumn{2}{c|}{$6d_{3/2}$} & \multicolumn{2}{c}{$6d_{5/2}$} \\ 
&  &   $A_h$  & $A_h$  & $B_h$  & $A_h$ & $B_h$ \\
\cline{3-7}
$^{223}$Ra$^+$  & &  &  & &  &  \\
& RCC & 3567.26 & 77.08 & 383.88 & $-$23.90 & 477.09 \\
& Expt. & 3404.0(1.9) & &        &          &        \\
$^{225}$Ra$^+$  & &  &  & &  &  \\
& RCC & $-$28977.76 & $-$626.13 & & 194.15 & \\
& Expt. & $-$27731(13) &   &     &      &      \\
\end{tabular}
\end{center}
\end{ruledtabular}
\caption{Hyperfine structure constants in MHz.}
\label{tab4}
\end{table}

The trends of the correlation effects in the hyperfine interactions of the $7s$ and $6d$ states in the present system are
similar to the corresponding states in Ba$^+$ \cite{bijaya2,bijaya6}. We have found 23\%, 31\% and 181\% correlation
contributions with respect to the DF results of $A_h$ in the $7s$, $6d_{3/2}$ 
and $6d_{5/2}$ states, respectively. The core-polarization (CP) effect in the $6d_{5/2}$ 
state is very strong and its contribution is larger than the DF result. This
gives rise to the unusual behavior of the electron correlation effects.

{\it Conclusion:} We have successfully carried out accurate calculations of the lifetimes, polarizabilities,
quadrupole moments and hyperfine structure constants in Ra$^+$ using the 
RCC theory. Our calculated values of the lifetimes of the  $6d$ states which 
are 0.893s and 0.297s, respectively, suggest that Ra$^+$ could be a suitable 
candidate for an optical frequency standard with an accuracy better than $10^{-18}$. The results of
the different properties that we have calculated can serve as benchmarks to 
guide experimentalists. On
the other hand, precise measurements of these quantities can also be used to 
test our method of calculation. 

{\it Acknowledgment:} We thank Dr. Manas Mukherjee for fruitful discussions.
These computations were performed on C-DAC's ParamPadma.

\end{document}